\documentclass[12pt,psfig]{article}
\usepackage{graphicx}
\begin{document}

\begin{center} 
{\bf  Effect of $\Lambda$ hyperons on the nuclear equation of state in a 
Dirac-Brueckner-Hartee-Fock model. 
 }              
\end{center}
\vspace{0.1cm} 
\begin{center} 
 F. Sammarruca \\ 
\vspace{0.2cm} 
 Physics Department, University of Idaho, Moscow, ID 83844, U.S.A   
\end{center} 
\begin{abstract}
We predict the energy per baryon in nuclear matter with 
non-zero fractions of $\Lambda$ hyperons. We include Dirac effects on the 
nucleon as well as 
the $\Lambda$ and describe how the latter are implemented. 
We use the 
  nucleon-hyperon meson-exchange potentials from the                    
 J{\"u}lich group, the latest as well as 
an earlier version. 
The dependence of the results on the many-body
framework and on the nucleon-hyperon interaction
model is discussed.   
\\ \\ 
PACS number(s): 21.65.+f,21.80.+a 
\end{abstract}

\section{Introduction} 
                                                                     
There are important motivations for including strange baryons in nuclear matter.
The presence of hyperons in the interior of neutron stars is reported to 
soften the equation of state (EoS), with the consequence that the predicted             
neutron star maximum masses become considerably smaller \cite{Sch+06}. With recent 
constraints allowing maximum masses larger than previously accepted limits 
\cite{Fuchs}, 
accurate microscopic calculations which include strangeness (in addition to other  
effects, such as those originating from relativity), 
become especially important and timely. 

Moreover, as far as terrestrial nuclear physics is concerned, studies 
of hyperon energies in nuclear matter naturally complement our knowledge
of hypernuclei (see, for instance, Refs.~\cite{hyper1,hyper2,hyper3,hyper4}.
For example, the EoS of hypermatter is 
useful in the development of generalized mass formulas depending on density and hyperon fraction \cite{catania1,hyper4}. 
From the experimental side, additional data are very much needed, especially
on $\Lambda \Lambda$ hypernuclei, which would provide information on the 
$\Lambda- \Lambda$ interaction. Concerning single hypernuclei, analyses of              
data on $\Lambda$ binding energies constrain 
the depth of the single-$\Lambda$ potential to be 27-30 MeV \cite{Dover}. 
The status of $\Sigma$ hypernuclei and the $\Sigma$-nucleus potential is more controversial 
(see Ref.~\cite{Saha04} and references therein).                 

Microscopic calculations of nuclear matter properties including hyperons have been reported earlier  
within the non-relativistic Brueckner-Hartree-Fock framework (BHF) (see, for instance, 
Refs.~\cite{catania1,catania2}), using             
the Nijmegen \cite{Nij89}         
nucleon-hyperon (NY) meson-exchange potential.                                     
Extensive microscopic work on hyperonic nuclear matter, again within the non-relativistic BHF framework, has also been published by 
the Barcelona group (see, for instance, Refs.~\cite{Vid+00,Vid+00b,Vid+04,Rios+05}). 

It is one purpose of the present work to bring in the new aspect of Dirac effects 
on the $\Lambda$ hyperon as well as the nucleon. By ``Dirac effects" we mean that the single-baryon            
 wavefunction is calculated self-consistently with the appropriate effective interaction.
The importance of these effects on the nucleonic equation of state cannot be overestimated \cite{Sam08b}.
They can be seen both as relativistic effects                                   
and three-body forces originating from nucleon-antinucleon excitations, and provide an essential saturation
mechanism missing from conventional approaches. 

Our previous calculation \cite{Sam08} of the binding energy of a 
$\Lambda$ impurity in nuclear matter showed that Dirac effects on the $\Lambda$ hyperon               
yield a moderate reduction of the binding energy. In that calculation, we used the most recent
nucleon-hyperon (NY) potential reported in Ref.~\cite{NY05} (thereafter referred to as NY05), and observed
that $G$-matrix predictions obtained with NY05 are                                  
significantly differerent from calculations using the previous 
(energy independent) version 
of the J{\"u}lich NY potential \cite{NY94}.                                                              
Therefore, in this work we will use both potentials, for comparison.                       
The Bonn B potential \cite{Mac89} is used throughout for the nucleon-nucleon (NN) part.     

Previous calculations of the EoS with hyperons have typically been conducted within a   
non-relativistic framework together with r-space local NY potentials. Our approach uses     
non-local relativistic momentum-space (NN and NY) potentials and a relativistic many-body 
method, 
and is therefore fundamentally different.                          

The baseline work for our EoS's was developed in Ref.~\cite{AS03}. The framework described there
for isospin-asymmetric nuclear matter has been further expanded and adjusted to include other species of baryons
in different concentrations. 
Our latest EoSs for the NN sector can be found in 
Ref.~\cite{SL08}.                                                                                         

In the next Section, we describe some technical aspects of the calculation. We then proceed to 
present and discuss our results (Section 3). Conclusions and future plans are summarized in Section 4. 

\section{Description of the calculations} 
\subsection{General formalism } 
For a total density of baryons (nucleons and $\Lambda$'s), $\rho$, and some specified $\Lambda$ fraction, 
$Y_{\Lambda}$, the densities of each species are known and are related to their respective
Fermi momenta by 
\begin{equation}
\rho _N = \frac{1}{3 \pi ^2}(2)(k_F^N)^3, \;\;\;\;\;\;
\rho _{\Lambda} = \frac{1}{3 \pi ^2}(k_F^{\Lambda})^3.                  
\end{equation}

For a nucleon and a $\Lambda$ with momenta ${\vec k}_N$ and 
${\vec k}_{\Lambda}$, the total and relative momenta are  
\begin{equation}
{\vec P} = {\vec k}_N + {\vec k}_{\Lambda}, \;\;\;\;\;\;\;\;
{\vec k} =\frac{M_N{\vec k_{\Lambda}} -M_{\Lambda}{\vec k_N}}{M_N + M_{\Lambda}}=
\alpha {\vec k_{\Lambda}} -
\beta {\vec k_N}, 
\end{equation}
with $\alpha$ and $\beta$ equal to $\frac{M_N}{M_N + M_{\Lambda}}$ and 
 $\frac{M_{\Lambda}}{M_N + M_{\Lambda}}$, respectively. Thus,                             
\begin{equation}
{\vec k}_{\Lambda} =\beta {\vec P} + {\vec k}, \;\;\;\
{\vec k}_{N} =\alpha {\vec P} - {\vec k}.          
\end{equation}

Clearly, for the case of two nucleons we have
\begin{equation}
{\vec P} = {\vec k}_N + {\vec k}_{N}', \;\;\;\
{\vec k} =\frac{{\vec k_{N}} -{\vec k_N}'}{2}. 
\end{equation}
The single-nucleon and single-$\Lambda$ potentials are obtained as 
\begin{equation}
U_N({\vec k_N}) = U_{N\Lambda} ({\vec k_N}) + U_{NN}({\vec k_N}),                    
\end{equation}
and 
\begin{equation}
U_{\Lambda}({\vec k}_{\Lambda}) = U_{\Lambda N} ({\vec k}_{\Lambda}),                                  
\end{equation}
i.e., the $\Lambda \Lambda$ interaction is neglected.  
In the equations above, the various terms, $U_{B_1B_2}$, are the contributions to the potential 
of baryon $B_1$ from its interation with all baryons of type $B_2$. They are given by
\begin{equation}
 U_{N\Lambda} ({\vec k_N}) = \sum_{T,L,S,J} \frac{(2T+1)(2J+1)}{(2t_N+1)(2s_N+1)}                               
  \int _0 ^{k_F^{\Lambda}} G^{T,L,S,J}_{N \Lambda}(k({\vec k_N},{\vec k_{\Lambda}}),  
  P({\vec k_N},{\vec k_{\Lambda}})) d^3 k_{\Lambda}, 
\end{equation}
\begin{equation}
 U_{N N} ({\vec k_N}) =                                                                  
 \sum_{T,L,S,J} \frac{(2T+1)(2J+1)}{(2t_N+1)(2s_N+1)}                               
 \int _0 ^{k_F^{N}} G^{T,L,S,J}_{N N}(k({\vec k_N},{\vec k_{N}'}),  
 P({\vec k_N},{\vec k_{N}'})) d^3 k_{N}', 
\end{equation}
and 
\begin{equation}
 U_{\Lambda N} ({\vec k}_{\Lambda}) =                                                                            
 \sum_{T,L,S,J} \frac{(2T+1)(2J+1)}{(2t_{\Lambda}+1)(2s_{\Lambda}+1)}                               
 \int _0 ^{k_F^{N}} G^{T,L,S,J}_{\Lambda N}(k({\vec k_N},{\vec k_{\Lambda}}),  
  P({\vec k_N},{\vec k_{\Lambda}})) d^3 k_{N}, 
\end{equation}
where the channel isospin $T$ can be 0 or 1 for the NN case and is equal to 1/2 for the 
$N\Lambda$ case, and $s_i,t_i$ ($i=N,\Lambda$) are the spin and isospin of the nucleon or
$\Lambda$. 

Notice that      
\begin{equation}
\frac{U_{N\Lambda}}{U_{\Lambda N}} \approx \frac{\rho _{\Lambda}}{\rho _N}, 
\end{equation}
an approximation often used in mean-field approaches. 

The average potential energies of nucleons and $\Lambda$'s are determined from 
\begin{equation}
<U_N> =\frac{1}{\rho_N}\frac{1}{(2 \pi)^3}4\frac{1}{2}\int_0^{k_F^N} U_{N} ({\vec k_N}) 
dk^3_N,          
\end{equation}
and 
\begin{equation}
<U_{\Lambda}> =\frac{1}{\rho_{\Lambda}}\frac{1}{(2 \pi)^3}2\frac{1}{2}\int_0^{k_F^{\Lambda}} U_{\Lambda } ({\vec k}_{\Lambda})
dk^3_{\Lambda},          
\end{equation}
where the factors of 4 and 2 in Eqs.~(11) and Eq.~(12), respectively, account for protons and neutrons in both spin states
or $\Lambda$'s in both spin states.

Finally the average potential energy per baryon is obtained as
\begin{equation}
<U> = \frac{\rho _N <U_N> + \rho _{\Lambda}<U_{\Lambda}>}{\rho _{tot}},
\end{equation}
from which, together 
with a similar expression for the kinetic energy, one obtains the average energy per 
baryon. 

The N$\Lambda$ $G$-matrix is obtained from the Bethe-Goldstone equation
\begin{eqnarray} 
<N\Lambda|G_{N \Lambda}(E_0)|N\Lambda> & = & <N\Lambda|V|N\Lambda> \\ 
                                       &   & \mbox{} +\sum_{Y=\Lambda,\Sigma} <N\Lambda|V|NY> \frac{Q}{E_0 - E}
<NY|G_{N \Lambda}(E_0)|N\Lambda> ,  \nonumber
\end{eqnarray}
where $E_0$ and $E$ are the starting energy and the energy of the intermediate NY state,  
respectively, and $V$ is an energy-independent NY potential.

For two particles with masses $M_N$ and $M_{\Lambda}$ and Fermi momenta
$k_F^N$ and $k_F^{\Lambda}$, Pauli blocking requires               
\begin{equation}
Q({\vec k},{\vec P}) = 
                \left\{
\begin{array}{l l}
 1          & \quad \mbox{$|\beta {\vec P}+{\vec k}|>k_F^{\Lambda}$  and 
	   $|\alpha {\vec P}-{\vec k}|>k_F^{N}$ }    \\            
 0          & \quad \mbox{otherwise.}          
\end{array}
\right.
\end{equation} 
The above condition implies the restriction 
\begin{equation}
\frac{(\frac{M_N}{M}P)^2 +k^2-(k_F^N)^2}{2Pk\frac{M_N}{M}} > cos \theta >
-\frac{(\frac{M_{\Lambda}}{M}P)^2 +k^2-(k_F^{\Lambda})^2}{2Pk\frac{M_{\Lambda}}{M}},                   
\end{equation} 
where $\theta$ is the angle between the total (${\vec P}$) and the relative
(${\vec k}$) momenta of the two particles, and $M=M_{\Lambda} + M_N$. Angle-averaging is then applied in the 
usual way. 

In the present calculation we consider a non-vanishing density of $\Lambda$'s but 
do not allow for the presence of real $\Sigma$'s in the medium 
(although both $\Lambda$ and $\Sigma$                        
are included in the coupled-channel calculation of the NY $G$-matrix, see Eq.~(14)).   
Essentially we are considering a scenario where a small fraction of nucleons is replaced
with $\Lambda$'s, as could be accomplished by an experiment aimed at producing multi-$\Lambda$ hypernuclei.
Multistrange systems, such as those produced in heavy-ion collisions, may of course contain other hyperons 
on the outset. 
For small $\Lambda$ densities, though, as those we consider here, the cascade ($\Xi$) and the $\Sigma$ hyperon
are expected to decay quickly through the strong processes 
 $N+\Xi \rightarrow \Lambda+\Lambda$ and 
 $N+\Sigma \rightarrow N+\Lambda$. Under these conditions, a mixture of nucleons and $\Lambda$'s 
can be considered ``metastable", in the sense of being equilibrated 
over a time scale which is long relative to strong processes  
\cite{catania1}. (In fact, the strong reactions mentioned above would have to be 
Pauli blocked in order to produce a metastable multistrange system \cite{SG00}.) 

We neglect the $YY'$ interaction, as very little is known        
about it. Furthermore, non-local momentum-space $YY$ potentials, appropriate for our DBHF framework, 
are not available at this time.  
For these reasons, we keep the $\Lambda$ concentration           
relatively low.

\subsection{Dirac effects on the $N\Lambda$ potential}                                

The relation between the non-relativistic Brueckner approach and the relativistic
framework (known as Dirac-Brueckner-Hartree-Fock, DBHF)                                 
has been discussed for a long time. Already in Ref.~\cite{Brown} it was
shown how relativistic effects tie in with virtual excitations of pair terms. 
Lately, these concepts have been revisited \cite{lombardo} with similar
conclusions. 
In short, 
the Dirac effect on the EoS of nucleonic matter is an essential saturating, and strongly  
density dependent, mechanism, which effectively accounts for the class of three-body forces
originating from virtual nucleon-antinucleon excitations. 
When hyperon degrees of freedom are included, for reasons of consistency, those should then be 
subjected to the same correction.            

We have incorporated DBHF effects in the $\Lambda$ matter calculation, which amounts to involving the 
$\Lambda$ single-particle Dirac wave function in the self-consistent calculation through the 
$\Lambda$ effective mass, $M^*_{\Lambda}$.                             
Similarly to what is done for nucleons in Ref.~\cite{Mac89}, we fit the single-particle energy for 
$\Lambda$'s using the ansatz
\begin{equation}
e_{\Lambda}(p) = \sqrt{(M^*_{\Lambda})^2 + p^2} + U^{\Lambda}_V, 
\end{equation}
where $M^*_{\Lambda}=M_{\Lambda} + U_S^{\Lambda}$, and 
$U_S^{\Lambda}$ and 
$U_V^{\Lambda}$ are the scalar and vector potentials of the $\Lambda$ baryon, which we 
assume to be momentum independent. Note the single-particle energy is the sum of the 
single-particle potential, for which we use the two-parameter ansatz, 
\begin{equation}
U_{\Lambda}(p) = \frac{M^*}{\sqrt{(M^*_{\Lambda})^2 + p^2}}U^{\Lambda}_S + U^{\Lambda}_V, 
\end{equation}
and the expectation value of the free energy operator in the Dirac equation \cite{Mac89}. 
Because of the small hyperon concentrations considered here, 
the parameters of the single nucleon potential are taken from a separate calculation of
symmetric nuclear matter \cite{AS03,SL08} performed at densities corresponding to the nucleon densities,
$\rho_N$ in Eq.~(1).
 This amounts to saying that, for low hyperon concentrations, the main correction to the 
single nucleon potential comes from the presence of the first term in Eq.(5). 
The parameters of the single $\Lambda$ potential are fitted, self-consistently with the $N\Lambda$
$G$-matrix, at momenta close to
its Fermi surface, specifically $k_1=0.8 k_F^{\Lambda}$ and 
$k_2= k_F^{\Lambda}$.         

A problem with the        
J{\"u}lich NY potential in conjunction with DBHF calculations                             
is the use of the pseudoscalar coupling for the interactions of 
pseudoscalar mesons (pions and kaons) with nucleons and hyperons. For the reasons 
mentioned above (that is, the close relationship between Dirac effects and ``Z-diagram"
contributions), 
this relativistic correction is known to become unreasonably large when applied to 
a vertex involving pseudoscalar coupling. On the other hand, the gradient (pseudovector) 
coupling (also supported by chiral symmetry arguments) largely suppresses antiparticle contributions. 
To resolve this problem, one can make use of the on-shell equivalence between the pseudoscalar and
the pseudovector coupling, which amounts to 
relating the coupling constants as follows: 
\begin{equation}
g_{ps} = f_{pv}\frac{M_i + M_j}{m_{ps}},                   
\end{equation} 
where $g_{ps}$ denotes the pseudoscalar coupling constant and $f_{pv}$ the pseudovector one;
$m_{ps}$, $M_{i}$, and $M_{j}$ are the masses of the                         
pseudoscalar meson and the two baryons involved
in the vertex, respectively. 
This procedure can be made plausible by                                                               
writing down the appropriate 
one-boson-exchange amplitudes and observing that, redefining the coupling constants as above, we have 
(see Ref.~\cite{Mac89} for the two-nucleon case)                  
\begin{equation}
V_{pv} = V_{ps} + .....             
\end{equation} 
where the ellipsis stands for off-shell contributions. 
 Thus, the pseudoscalar coupling can be interpreted as pseudovector coupling where the 
off-shell terms are ignored. This is what we apply in our DBHF calculations.                  

In the coupled channel calculation, evaluation of $G_{N \Lambda}$ involves 
the transition potentials $V_{N \Lambda\rightarrow N \Lambda}$,
$V_{N \Lambda \leftrightarrow N \Sigma}$, and 
$V_{N \Sigma \rightarrow N \Sigma}$                                                   
(all with total channel isospin equal to 1/2), plus the corresponding exchange diagrams. 
Because in the present scenario                                                                                     
the $\Sigma$ hyperon is not given an effective mass,               
Dirac effects are applied         
only in $V_{N \Lambda \rightarrow N \Lambda}$. 
A diagram where not all of the baryon lines are Dirac-modified may yield a Dirac effect that is
artificially skewed.                                                                                             

Finally, a comment is in place concerning meson propagators. In standard DBHF calculations \cite{Mac89},
the so-called Thompson equation (a relativistic three-dimensional reduction of the 
Bethe-Salpeter equation) is used for two-baryon scattering. In the Thompson formalism, static
propagators           
are employed for meson exchange, i.e. 
\begin{equation}
-\frac{1}{m_{\alpha}^2 + ({\vec q}' - {\vec q})^2}  
\end{equation} 
where $m_{\alpha}$ denotes the mass of the exchanged meson and ${\vec q},{\vec q'}$ are 
the baryon momenta in their center-of-mass frame before and after scattering. 
The J{\"u}lich NY potentials are 
based upon time-ordered perturbation theory \cite{NY89} and use a meson propagator
given by 
\begin{equation}
\frac{1}{\omega_{\alpha} (z-E_i -E_j - \omega _{\alpha})} 
\end{equation} 
with $\omega _{\alpha}=\sqrt{m_{\alpha}^2 + ({\vec q'}-{\vec q})^2}$; 
$E_i=\sqrt{M_i^2 + {\vec q}'^2}$ and 
$E_j=\sqrt{M_j^2 + {\vec q}^2}$ are baryon energies and $z$ is the starting energy of the two-baryon 
system. In order to eliminate the energy dependence, Reuber {\it et al.} \cite{NY94} replaced the original
$z$ with 
\begin{equation}
z = \frac{1}{2} (M_1 + M_2 + M_3 + M_4),                                  
\end{equation} 
where the $M_i$'s denote the baryon masses of the four legs in the one-meson exchange diagram.
In any case, the J{\"u}lich meson propagator involves the baryon masses. Replacement of these
free-space masses with in-medium values would create medium effects on meson propagation which we do 
not wish to include in our nuclear matter calculations. The reason for keeping free-space masses in the 
meson propagator is twofold. First, standard DBHF calculations do not include medium effects
on meson propagation as they typically use Eq.~(19), which does         
not depend on baryon masses. Second, medium effects on meson propagation constitute a separate class of 
effects that we are not concerned with in the present context  
and are typically not perceived as part of the DBHF approach.

Having taken the steps described above, 
we proceed to the DBHF calculations. The effect of such corrections on the binding energy of a $\Lambda$
impurity in otherwise pure nuclear matter was reported 
in Ref.\cite{Sam08}. 

\section{The equation of state of nucleon and $\Lambda$ matter}                                 

As stated in the Introduction, 
we will be showing results for both the NY05 potential \cite{NY05} and the previous version of the 
J{\"u}lich NY potential, NY94, specifically the model referred to as ${\tilde A}$ 
in Ref.~\cite{NY94}.

In Fig.~1 we show the energy per particle as a function of density for different $\Lambda$
concentrations as obtained from DBHF calculations along with the NY94 potential.                 
As more nucleons are replaced with 
$\Lambda$'s, generally less binding energy per particle is generated. This is due to the weaker nature of the
$N\Lambda$ interaction relative to the NN one. 
Furthermore, the $\Lambda$ Fermi momentum grows rather quickly with $\rho _{\Lambda}$, since only
two $\Lambda$'s can occupy each state, rather than four, which implies a fast rise of the 
hyperon kinetic energy. (Although, for small hyperon densities,                  
there is at first some reduction of the kinetic energy, due to the fact that          
$\Lambda$'s have larger mass and can occupy lower energy levels.)     
We notice that the saturation density remains essentially unchanged 
with increasing hyperon concentrations. 
As density grows, though, larger $\Lambda$ concentrations start to yield increased attraction.
One must keep in mind that, especially at the higher densities, the NN interaction become less and less
attractive due to medium effects, in particular repulsive Dirac effects. Thus, removing nucleons
from the system can actually amount to increased binding.

Moving now to Fig.~2, where the NY05 potential is adopted,                           
we see a very different scenario.                                    
We recall that the NY05 model is considerably more attractive, yielding
about 50 MeV for the $\Lambda$ binding energy \cite{Pol,Sam08} whereas a value close to 30 MeV was found with NY94         
\cite{NY94}. 
Naturally, we expect these differences to reflect onto the respective EoS predictions.
Here, increasing the hyperon population yields more binding, 
a trend opposite to the one seen in the previous figure.
Again, the final balance is the result of a combination of effects.
The fact that the NN component is repulsive at the higher densities, 
together with the more attractive nature of the NY05 potential, determines here a net increase in attraction 
with decreasing {\it nucleon} density. 
However, the additional binding becomes smaller and
smaller with increasing hyperon concentration, indicating that, at sufficiently large $\Lambda$
densities, the net balance may turn repulsive.             
Notice also that in Fig.~2 the minimum moves towards higher densities, signifying that
baryon pairs favor a smaller interparticle distance.

To summarize, one must keep in mind that the energy/particle is the result of a delicate balance          
of both the kinetic energies and the potential energies of the two baryon species, being weighed by
the respective densities. 
Thus, although the NN interaction is generally more attractive than the N$\Lambda$ one, 
the net effect of replacing  nucleons with $\Lambda$'s will depend 
sensitively on the nature of the NY interaction that's being put into the system, as well as the 
``stiffness" of the original, nucleonic, EoS. 

This is further confirmed in 
Fig.~3, where we show predictions from conventional BHF calculations (i.e., no ``Dirac" effects). 
As in the previous figures, 
the red curve is the nucleonic EoS. The solid(dashed) curves are the predictions with NY94(NY05).  
Clearly, the effects are opposite depending on the NY interaction. Qualitatively, the trend seen
in each group of curves (solid or dash) is approximately consistent with the one observed previously in the
corresponding (i.~e. same NY potential) DBHF predictions. However, the ``cross over" of the curves                       
seen in Fig.~1 at about twice saturation density is due to a large extent to the more repulsive nature of the
nucleonic EoS in the DBHF calculation (see comments above). 

In conclusion, the effect of hyperons on the EoS is strongly dependent upon the baseline
(nucleonic) EoS as well as the NY potential model.                                              
With regard to the first issue, we stress the importance of starting from a realistic EoS, such as the one
predicted in DBHF calculations or obtained with the inclusion of three-body forces. 
The second observation confirms the conclusions of Ref.~\cite{vlowk}                
concerning the large uncertainties originating from the bare NY potentials.
Unfortunately, the existing data do not set sufficient constrains on the potentials, as demonstrated by
the fact that different potentials can fit the available scattering data equally accurately but
produce very different scattering lengths \cite{vlowk}.

\begin{figure}
\begin{center}
\vspace*{-4.0cm}
\hspace*{-2.0cm}
\scalebox{0.5}{\includegraphics{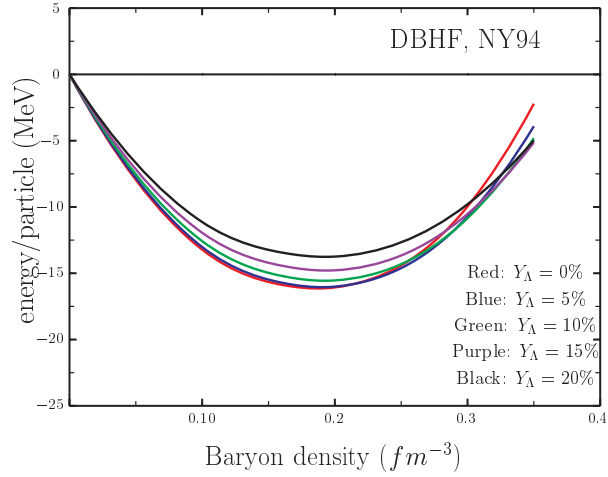}}
\vspace*{-2.5cm}
\caption{(Color online) Energy/particle as a function of density in symmetric nuclear matter for various
$\Lambda$ concentrations $Y_{\Lambda}$. Predictions obtained from DBHF calculations with the NY94 potential.
} 
\label{one}
\end{center}
\end{figure}
\begin{figure}
\begin{center}
\vspace*{-4.0cm}
\hspace*{-2.0cm}
\scalebox{0.5}{\includegraphics{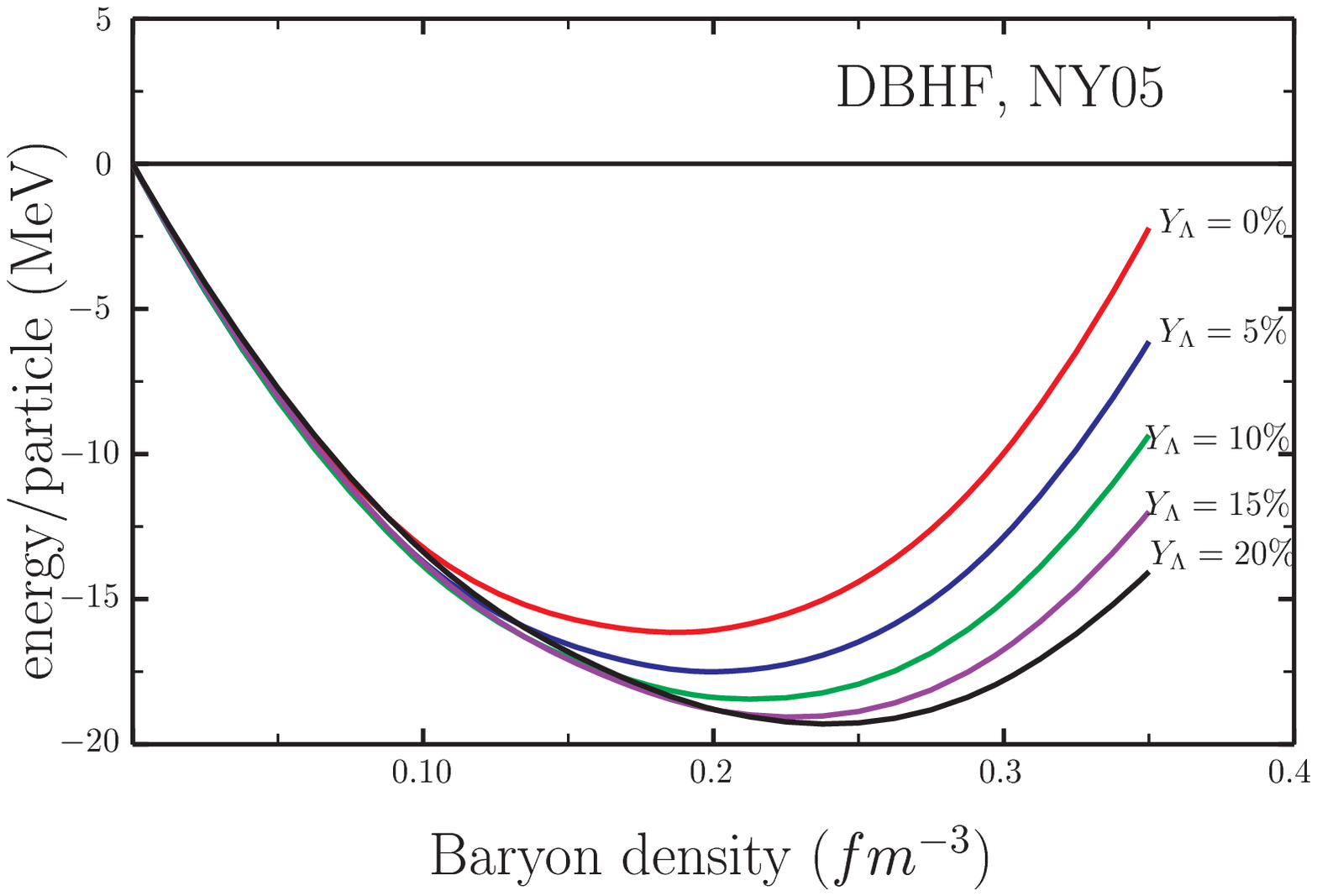}}
\vspace*{-3.0cm}
\caption{(Color online) Energy/particle as a function of density in symmetric nuclear matter for various
$\Lambda$ concentrations $Y_{\Lambda}$. Predictions obtained from DBHF calculations with the NY05 potential. 
} 
\label{two}
\end{center}
\end{figure}

\begin{figure}
\begin{center}
\vspace*{-4.0cm}
\hspace*{-2.0cm}
\scalebox{0.5}{\includegraphics{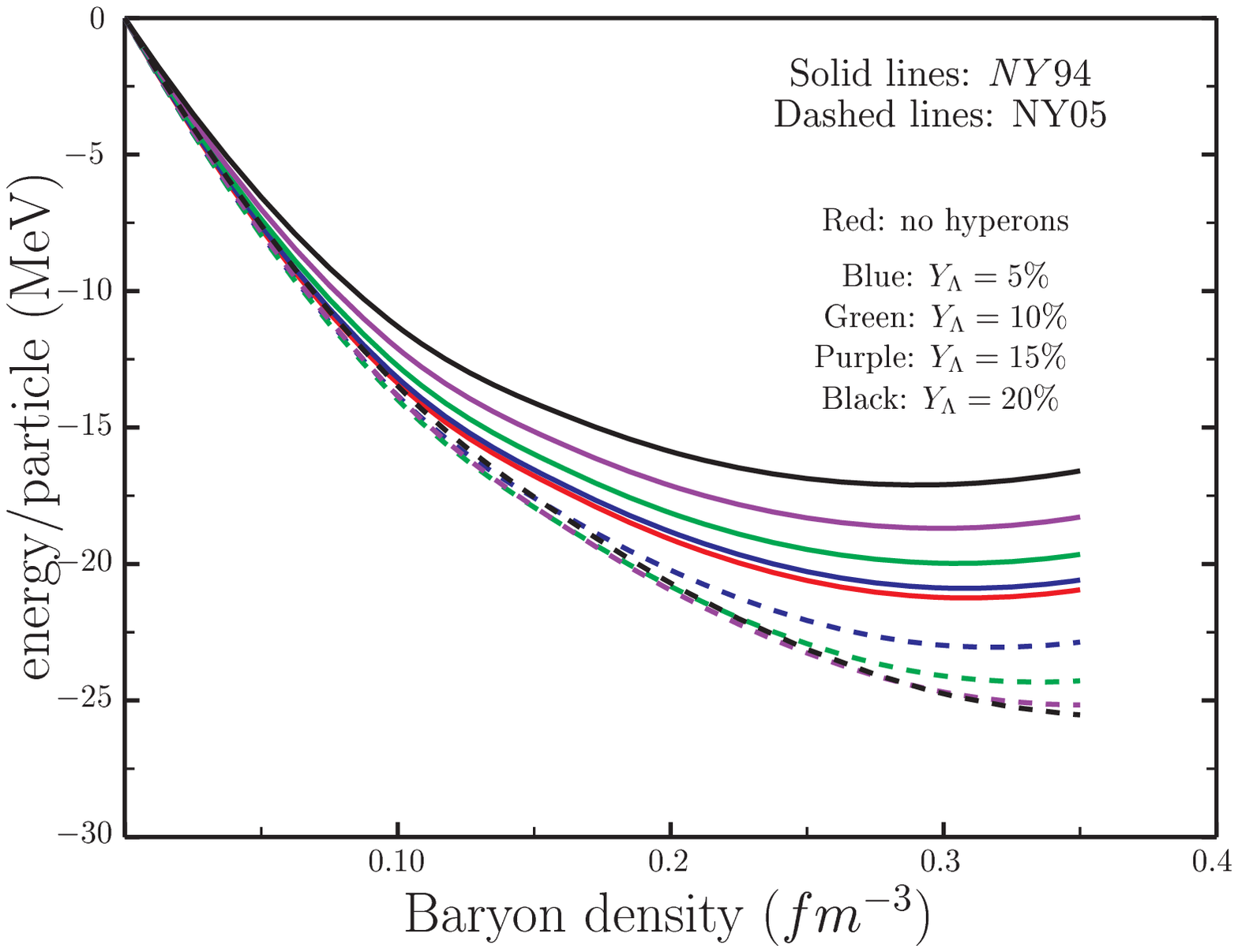}}
\vspace*{-2.0cm}
\caption{(Color online) Energy/particle as a function of density in symmetric nuclear matter for various
$\Lambda$ concentrations. Predictions obtained from BHF calculations. The solid(dashed) lines
are obtained with the NY94(NY05) potentials. 
} 
\label{three}
\end{center}
\end{figure}

\section{Conclusions }                                
We have reported  on                               
Dirac-Brueckner-Hartree-Fock predictions of the energy per particle in symmetric nuclear matter 
as a function of the total baryon density and (moderate) $\Lambda$ concentrations. 
Dirac effects are included on both the NN and the N$\Lambda$ potentials. Our 
DBHF scheme, which requires the use of relativistic momentum-space nucleon-baryon potentials,
represents a fundamentally different paradigm as compared to existing calculations.

Ultimately,
the actual fraction of hyperons present in star matter must be determined by the equations
of $\beta$ stability and charge neutrality for highly asymmetric matter containing neutrons, protons, hyperons,
and leptons.
Unlike the straightforward calculation of lepton fractions in $\beta$-stable isospin-asymmetric nucleon matter, 
the case of nucleons and hyperons require a lot more effort due to their strongly interacting nature.
One needs to know the chemical potentials of
each species (neutrons, protons, various hyperons) ideally at any densities and baryon
concentrations. Those may then have to be fit with analytic functions (not a trivial task, given
the number of dependences), so that they 
can be inserted into the appropriate equations 
to determine the equilibrium fractions of each baryon type. 
This is deferred to another paper. 

We observed that the predicted energy/particle is extremely sensitive to the 
chosen NY interaction.                                                                   
The large uncertainties due to the model dependence discussed in this 
paper are likely to impact any conclusions on the properties of strange neutron stars,
which therefore must be interpreted with caution. These include considerations of deconfinement
and possible transition from hadronic to quark matter, which depend sensitively on the equation 
of state in the hadronic phase.

\section*{Acknowledgments}
Support from the U.S. Department of Energy under Grant No. DE-FG02-03ER41270 is 
acknowledged. I am very grateful to Johann Haidenbauer for providing the nucleon-hyperon
potential codes and for useful communications. 

\end{document}